\documentstyle{article}

\begin{document}
\baselineskip = 20pt
\input epsf

\parindent 25pt
\overfullrule=0pt
\tolerance=10000
\def\Re{\rm Re}
\def\Im{\rm Im}
\def\titlestyle#1{\par\begingroup \interlinepenalty=9999
     \fourteenpoint
   \noindent #1\par\endgroup }
\def\tr{{\rm tr}}
\def\Tr{{\rm Tr}}
\def\half{{\textstyle {1 \over 2}}}
\def\quart{{\textstyle {1 \over 4}}}
\def\calt{{\cal T}}
\def\ie{{\it i.e.}}
\def\np{Nucl. Phys.}
\def\pl{Phys. Lett.}
\def\pr{Phys. Rev.}
\def\prl{Phys. Rev. Lett.}
\def\cmp{Comm. Math. Phys.}
\def\quart{{\textstyle {1 \over 4}}}
\def\RR{${\rm R}\otimes{\rm R}~$}
\def\NSNS{${\rm NS}\otimes{\rm NS}~$}
\def\RNS{${\rm R}\otimes{\rm NS}~$}
\def\calf{${\cal F}$}
\def\be{\begin{equation}}
\def\ee{\end{equation}}

\baselineskip=14pt

\thispagestyle{empty}

{\hfill RU-97-33 }

{\hfill  CPTH/S 506.0597}

{\hfill DAMTP/97-45}

\vskip 0.1cm
{\hfill hep-th/9705074}
\vskip 2.5 cm

\begin{center}
{\large\bf ANOMALOUS CREATION OF BRANES} \break

 \centerline{ Constantin P. Bachas$\ ^a$,   Michael R.  Douglas$\ ^b$  
and Michael
B.
Green$\ ^c$}

\vskip  0.5in
\centerline{$\ ^a$\  Centre de Physique Th\'eorique,  Ecole
Polytechnique,}
\centerline{ 91128 Palaiseau, France}
\centerline{bachas@pth.polytechnique.fr}
 \vskip 0.3cm

\centerline{$\ ^b$ Department  of Physics and Astronomy}
\centerline{ Rutgers University,  Piscataway, NJ
08855-0849, USA}
\centerline{
mrd@physics.rutgers.edu }
\vskip 0.3cm

\centerline{$\ ^c$ Department of Applied Mathematics and
Theoretical Physics,}
\centerline{Silver Street,Cambridge CB3 9EW, United Kingdom}
\centerline{M.B.Green@damtp.cam.ac.uk}

\end{center}
\vskip 1.6 cm
\rm

\begin{quote}
{\bf ABSTRACT:}
 In certain circumstances when two branes pass through each other a
third brane
is produced stretching between them.  We explain this phenomenon
by the use of
chains of dualities and the  inflow of charge that is required for
the absence
of chiral gauge anomalies when pairs of D-branes intersect.

\end{quote}
\vskip1cm

\normalsize

\newpage
\pagestyle{plain}
\setcounter{page}{1}

%\section{Introduction}

\def\baselinestretch{1.2}
\baselineskip 16 pt
\noindent
 \setcounter{equation}{0}

 At large distances D-branes \cite{polchina, J2} interact through the
long-range
 fields of supergravity.  At sub-stringy scales, on the other hand,
 their interaction is more usefully described in terms of  the
creation and
 subsequent annihilation of  (virtual) pairs of
 open strings stretching between the branes. Such pairs may
 materialize from the vacuum \cite{bachasa}, a phenomenon analogous
 to pair creation in an electric field.  There are however  special
 situations, to be discussed here,  in which {\it a single} open
string {\it
must}  materialize
 when two D-branes cross.
  Hanany and Witten \cite{HW}    pointed out a similar
phenomenon,
in which a stretched
 D3-brane is produced  when a D5-brane and a NS 5-brane cross. In
this note we
 will explain how such phenomena can be related by chains of dualities
 to the (abelian) anomaly equation describing charge inflow
\cite{callan} on
the intersection of branes \cite{GHM}.

A set of $m$ coincident  type-II D-branes is described by a
supersymmetric
field theory with gauge group $U(m)$. When two such sets of D-branes
intersect, the field content in the intersection domain
includes the pull-backs of the corresponding
$U(m)$ and $\tilde U({\tilde m})$ gauge potentials,
as well as  the pull-back of the bulk graviton field.
There are circumstances in which this
effective field theory  is chiral \cite{GHM}.  For this to
arise   the
two sets of coincident branes must  intersect
in such a manner that  the space-time
dimension of the intersection domain is  $2$ mod $4$, and there
are  no spatial
directions  transverse to both sets.
Two distinct  configurations of intersecting type-II D-branes that
result in
chiral supersymmetry in the intersection region are: {\it (a)}
 two D5-branes of type-IIB theory intersecting on a
string, or {\it (b)} two D7-branes of type-IIB theory
intersecting on a 5-brane.
Other chiral configurations  in either type IIA or IIB are obtained
from these
by T-dualizing coordinates transverse to the intersection region.
Let us denote  the
world-volume actions of  the two individual branes  and of the
intersection region by ${\cal S}$, $\tilde {\cal S}$
and  ${\cal I} $, respectively.
Under a gauge transformation,
the variation of ${\cal S}$ and $\tilde {\cal S}$
has a (boundary) piece localized at the  intersection, which precisely
cancels  the anomalous variation of ${\cal I}$
 \cite{GHM}.   Physically, since the embedding theory as well as
the theory on
either of   the two  D-branes  is non-anomalous,
the apparent charge violation in the presence of background fields
in   the intersection region
has to  be  accounted for by  inflow of charge from the  branes.
This is an
application of the `anomaly inflow' argument of \cite{callan}.

 Let us focus  first on  configuration  (a)  with
two intersecting D5-branes lying in the planes $(12345)$ and
$(16789)$, respectively.
We will  take the direction $X_1$ to be a circle
with radius $L_1$. Furthermore, it will be sufficient to
consider the abelian case, in which
there is an independent  maximally-supersymmetric $U(1)$ vector
multiplet
living in each
brane.  The spectrum in the
 intersection  region consists of  the pull-backs of these two
abelian gauge
potentials,
  together with a single  Weyl fermion  with $U(1)\times {\tilde U(1)}$
charge  $(+, -)$.  This fermion, the lightest state of an
oriented open string
stretching between the two intersecting branes,
does not transform under $(0,8)$ supersymmetry.
 Now consider  switching
 on  time-dependent Wilson lines  $A_1(t)$
and ${\tilde A}_1(t)$ on the two branes.
 The anomaly equation implies that the
net inflow of fermion number onto the intersection domain is
\be
\Delta N =  L_1 \int  dt\ \left({d A_1\over dt} -
 {d {\tilde A_1}\over dt}\right) .
\ee
 A  T-duality  along the first dimension transforms
 the D5-branes into  D4-branes
parallel to the hyperplanes $(2345)$ and $(6789)$. Furthermore
 the gauge fields
$2\pi\alpha^\prime A_1$ and $2\pi\alpha^\prime
{\tilde A_1}$ become the  transverse
coordinates $X_1$ and ${\tilde X}_1$ of the 4-branes,
 while  the radius $L_1$ gets mapped  to
$L_1^\prime= \alpha^\prime / L_1$. The anomaly equation thus
takes the form
\be
\Delta N = {1\over 2\pi  L_1^\prime}
 \int  dt\ \left( {d X_1\over dt} - {d {\tilde X}_1 \over dt}\right) \ ,
\ee
where the  right-hand side may now be recognized as
the net (forward minus backward) number of times the two D4-branes cross
while moving on the circle of radius $L_1^\prime$.
This phenomenon can be understood as follows:  the energy levels of a
ground-state (chiral
fermionic) open string stretching between the two branes are
\be
 E_n =  {1\over 2\pi\alpha^\prime}  (X_1 - {\tilde X}_1 + 2\pi n
L_1^\prime ),
\ee
where $n$ is the number of times the string winds around the circle.
 Every time  the two D4-branes pass through each other,
an  energy level crosses the zero axis creating a
single   particle or hole.  This is precisely the content of the anomaly
equation, as illustrated in figure 1.

\begin{figure}
\centerline{\epsfbox{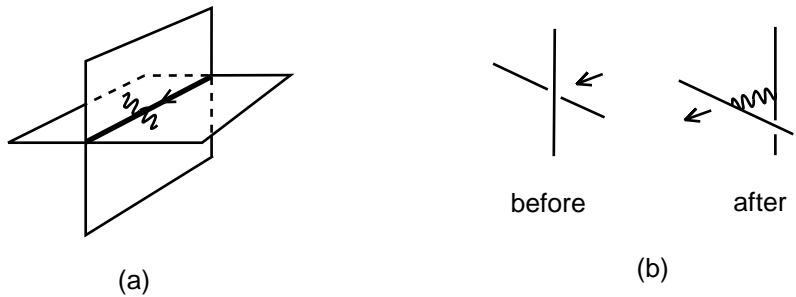}}
\caption{(a)  Two five-branes intersecting in a single compact direction.
Turning
on an electric  field along the circle induces an electric current
due to the
flow of open strings joining the branes.  (b)  After T-duality
this describes  two D4-branes that have a relative velocity along
the circle
and an open string is created each time the branes pass through
each other.}
\end{figure}

By using  chains of dualities  this process can be transformed into
 more exotic  phenomena. We will denote a process
  in which two branes, $A$ and $B$,  create a stretched
 brane $C$ when they pass through each other by
 $A\bigotimes B \hookrightarrow C$.   The  basic
example, the creation of a stretched fundamental string when two
 mutually transverse D4-branes cross,  reads in this notation:
 $ D(2345) \otimes D(6789) \hookrightarrow F(1)$.  The parentheses
  indicate the hyperplanes along which the various branes lie.
 Consider now the following sequence of dualities,
 \[
\begin{array}{ccc}
({\rm IIA})\ \ \ \ \  &\ \  D(2345) \bigotimes  D(6789)  \hookrightarrow
F(1) \\
 \ & \\
 \  & \Biggl\downarrow { T{(6)\ \ \ } }   \\
 \ & \\
({\rm IIB})\ \ \ \ \   & D(23456) \bigotimes   D(789)
\hookrightarrow F(1) \\
  \ & \\
\  & \Biggl\downarrow { S{\ \ \ \ \ \ } } \\
\ & \\
({\rm IIB})\ \ \ \ \  & NS(23456) \bigotimes  D(789)
\hookrightarrow  D(1) \\
\ & \\
 \  & \Biggl\downarrow { T(56) }  \\
 \ & \\
({\rm IIB})\ \ \ \ \   & \ \ \ \ \ \ NS(23456) \bigotimes D(56789)
\hookrightarrow D(156)
\end{array}
\]
\vskip 0.2cm
\noindent Here  $T(ij...)$ is a T-duality along the indicated dimensions,
while $S$  is the strong-weak coupling duality, which exchanges the
Neveu-Schwarz ($NS$)
and Dirichlet fivebranes of type IIB,
leaves the D3-brane invariant, and transforms
the fundamental string to a D-string.
What we learn from this chain of dualities is that a D3-brane
must be created when a NS-brane and a D5-brane, sharing two common
dimensions, pass through each other.
 This is precisely the phenomenon
observed by Hanany and Witten \cite{HW}.

Another sequence of dualities gives
\[
\begin{array}{ccc}
({\rm IIA})\ \ \ \ \  &\ \ \ \  D(2345) \bigotimes  D(6789)
\hookrightarrow
F(1) \\
 \ & \\
 \  & \Biggl\downarrow { T(67) }   \\
 \ & \\
({\rm IIA})\ \ \ \ \   & D(234567) \bigotimes   D(89)
\hookrightarrow F(1) \\
  \ & \\
\  & \Biggl\downarrow { S{\ \ \ \ \  } } \\
\ & \\
({\cal M}) \ \
 & KK(234567) \bigotimes  M(89)  \hookrightarrow  M(1\, 10) \\
\end{array}
\]
\vskip 0.2cm
\noindent where $S$ is now the duality that lifts type-IIA theory to
 ${\cal M}$-theory compactified on a circle. ${\cal S}$  transforms the
 D6-brane to the Kaluza-Klein monopole and  lifts    the D2-brane to the
M-brane  transverse to the tenth dimension.  It also lifts the
fundamental
string  to the {\it wrapped} M-brane.  We conclude  that a wrapped
M-brane must
be created whenever
 an unwrapped one passes through a transverse Kaluza-Klein D6-brane.
Alternatively,  the original pair of  crossing D4-branes can be lifted
directly
 to ${\cal M}$-theory where the process corresponds to  the creation of a
 wrapped membrane
whenever two wrapped but otherwise transverse fivebranes cross each
other.

Our discussion so far has been  based on the gauge anomaly equation in
two dimensions, which we have  interpreted  in terms of the net
number  of times
two D4-branes cross  when moving along their  unique common transverse
direction.
The   six-dimensional
anomaly equation in the overlap domain of two (sets of)
D7-branes (case
(b) above) can be given a similar interpretation. 
 The generic situation consists of  $m$  $D(1234567)$ and $\tilde m$
$D(1234589)$ branes,  intersecting
on some five-manifold ${\cal M}^5$ which spans the dimensions $(12345)$.
The
net inflow of fermions on this intersection domain reads \cite{GHM}
\be
\Delta N =  \int_{R \times{\cal M}^5}
\tr_m \ {\rm exp}\left( {{\cal F}\over 2 \pi} \right)\ \
\tr_{{\tilde m}^*} \ {\rm exp}\left( {{\tilde {\cal F}}\over 2 \pi}\right)
\ {\hat A}({\cal R} )\
\Biggl\arrowvert_{ 6-{\rm form}},
\ee
where ${\cal F}$ and $\tilde {\cal F}$ are the (pull backs) of the  
gauge field
 strengths and ${\cal R}$ is the (pull back) of the curvature.
 The traces are in the fundamental (conjugate)  representation of
$U(m)$ (respectively $\tilde U(\tilde m)$),
 and multiplication is in the sense of forms.
 The roof genus has an expansion in forms,
  $\hat A = 1 - p_1({\cal R})/24 + \cdots$, where $p_1$ is the
first Pontryagin class  and $\cdots$
represents higher-rank $4n$-forms.

We shall consider  the particular example in which the intersection  
domain is
the  orthogonal torus, ${\cal M}^5
 = (S^1)^5$, where  each of the five circles has radius  $L_i$
($i=1,\cdots,5$).
 It will also be  sufficient to consider $m$ coincident seven-branes
intersecting a single seven-brane ($\tilde m =1$)  in a five-dimensional
region, and to switch off the $U(1)$ gauge field
on the latter.
 The anomalous $U(1)$ charge is  induced by first turning on a
time-dependent Wilson line, $A_1(t)$,  along the $x_1$ direction in
the abelian $U(1)$ factor of the $U(m)$ gauge group, and taking
all other fields   to be independent of $t$ and $x_1$.
A T(1) duality transformation maps this configuration to
    a bunch of coincident six-branes,  $D(234567)$  moving
  around the dual circle of radius $L_1^\prime$  and passing through
  a single six-brane, $D(234589)$,  that is stationary.
The branes intersect instantaneously on the  four-torus $(S^1)^4$
that spans the directions ($2345$) and the anomaly  equation takes   
the form
\be
\Delta N =  {1\over 2\pi L_1^\prime} \int dt {dX_1\over dt} \times
 \int_{(S^1)^4}  \  {1\over 4\pi^2}\tr_m  {\cal F}
  \wedge  {\cal F}     ,
\ee
where
$X_1$ is the position on the (dual) circle
of the $m$ coincident moving branes.
The net charge inflow is thus given by
\be
\Delta N = n_{c} k \ ,
\ee
where $n_c$ is  the net number of times the two sets of branes  
encounter each
other and $k$ is the instanton number
 of the $U(m)$ gauge fields living on the moving branes  and
pulled back to the four-torus $(S^1)^4$.

We may interpret equation (6)
in terms of  the basic phenomenon  of crossing D4-branes, 
 described previously in figure 1.
 Indeed, in the limit where the $k$ gauge instantons have
(almost) zero size, they are equivalent to $k$ two-branes,
$D(67)$,  bound to the $m$ coincident moving six-branes
 \cite{Witt,Doug}. As these
 D2-branes pass through the stationary D6-brane they
each create
one fermionic string, a process T-dual to the process
of  figure 1.
An alternative gauge configuration consists of smooth Wilson lines
in the $U(1)$ factor of the  $U(m)$ gauge group,
\be
A_2 = { n_{2} x_4 \over 2\pi L_2 L_4 } , \qquad\qquad
A_3 = { n_{3}  x_5 \over 2\pi L_3 L_5 }.
\ee
Here   $n_{2}$ and $n_{3}$ are integers,
consistently with the quantization of magnetic flux, and
the total instanton number is
$k = n_{2} n_{3} m$.  A  T(23) duality transformation
maps all D6-branes
to D4-branes.  The stationary brane now  spans  the hyperplane  
$(4589)$  while
all $m$  moving branes are uniformly rotated away from
 the hyperplane $(4567)$ since the Wilson lines become the
 transverse  displacements
\be
X_2 = n_{2}  {L_2^\prime \over L_4} x_4, \qquad\qquad
X_3 = n_{3}   {L_3^\prime \over L_5} x_5 \ .
\ee
One may visualize the rotated branes as spiralling around the
(dual) torus $(234567)$. Each such spiralling
 D4-brane intersects  the stationary D4-brane in $n_2 n_3$ points
as it passes through, in accordance again with the
anomaly equation (6) (see figure 2).

The case ${\cal M}^5 = S^1\times K3$ can be analyzed similarly.
The T(1) duality transformation, which sets the branes in motion,
must  now be followed by a mirror map  which acts as T-duality
on the fiber tori of the $K3$ surface \cite{Ooguri}.
The end result is (sets of) D4-branes, which intersect the
$K3$ surface on two-cycles and intersect each other at points as they
cross.  The mirror map from these two-cycles
to marked points and/or the entire surface is  subtle, however,
because  the latter carries a negative unit of induced point-charge
\cite{Vafa}.

\begin{figure}
\centerline{\epsfbox{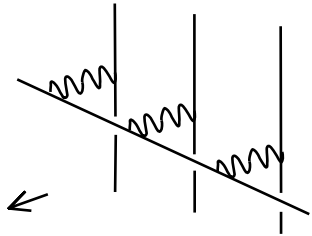}}
\caption{ Two sets of  D4-branes, arising  by T-duality from  intersecting
D7-branes, produce a number of strings equal to the total number of
intersection
points as they move past each other.
The horizontal line is the stationary brane,
whereas  the vertical lines represent either (a) $k$ zero-size
instantons, or (b)  $k$ segments of the $m$  
(continuous) spiralling and  moving
branes,  discussed in the text.
}
\end{figure}

\vskip 1cm

{\bf  Aknowledgements}

We thank T. Banks and P. Townsend for useful conversations.
C.B. gratefully aknowledges the hospitality of
the  high-energy group at Rutgers  and  of the
Newton Institute, during completion of this work. The research of
C.B. and M.B.G. was supported in part by the EEC grant
CHRX-CT93-0340.

\vfill\eject


\vspace{2cm}


\begin{thebibliography}{6666}

\bibitem{polchina} J. Polchinski, {\it Dirichlet Branes and Ramond-Ramond
Charges},   hep-th/9510017, Phys. Rev. Lett. {\bf 75} (1995) 4724.

\bibitem{J2} S. Chaudhuri, C. Johnson and J. Polchinski, {\it Notes on
D-Branes}, hep-th/9602052; J.  Polchinski, {\it TASI lectures on
D-branes},
hep-th/9611050.


\bibitem{bachasa} C. Bachas, {\it D-Brane Dynamics},     
hep-th/9511043,Phys.
Lett.
{\bf B374}
(1996)
37.

\bibitem{HW} A. Hanany and E. Witten, {\it Type-IIB Superstrings, BPS
Monopoles and Three-Dimensional Gauge Dynamics}  , hep-th/9611230.

\bibitem{callan} C.G. Callan and J.A.  Harvey,
 {\it Anomalies and Fermion Zero Modes on Strings and Domain Walls},
  Nucl. Phys. {\bf B250} (1985) 427.

\bibitem{GHM} M.B. Green, J.A.  Harvey and G. Moore, {\it I-Brane Inflow
and Anomalous Couplings on D-Branes}, hep-th/9605033,  Class.  
Quant. Grav. {\bf
14}   (1997) 47.

\bibitem{Witt} E. Witten, {\it Small Instantons in String Theory},
hep-th/9511030, Nucl. Phys. {\bf B460}  (1996) 541.

\bibitem{Doug} M.R. Douglas, {\it Branes within Branes}, hep-th/9512077.

\bibitem{Ooguri} H.  Ooguri, {\it Aspects of D-branes in Curved Space},
 Spring School on String Theory,  Gauge Theory and
  Quantum Gravity, ICTP, Trieste (April 1997).


\bibitem{Vafa} M. Bershadsky, C. Vafa and V. Sadov, {\it
D-branes and Topological Field Theories}, hep-th/9511222, Nucl.  
Phys. {\bf
B463}  (1996) 420.

\end{thebibliography}
\end{document}